\documentclass[epj]{webofc}
\usepackage[varg]{txfonts}   % Web of Conferences font
\woctitle{7th International Conference on Acoustic and Radio EeV Neutrino Detection Activities}
\begin{document}
\title{Measurement of horizontal air showers with the Auger\\ Engineering Radio Array}

\author{\firstname{Olga} \lastname{Kambeitz on behalf of the Pierre Auger Collaboration}\inst{1,2}\fnsep\thanks{\email{olga.kambeitz@kit.edu}} %\and
%        \firstname{Second author} \lastname{Second author}\inst{2}\fnsep\thanks{\email{Mail address for second
%             author if necessary}} \and
%        \firstname{Third author} \lastname{Third author}\inst{3}\fnsep\thanks{\email{Mail address for last
%             author if necessary}}
        % etc.
}

\institute{Institut f\"ur Kernphysik, \\Karlsruhe Institute of Technology (KIT), Germany
\and
Full author list: \url{http://auger.org/archive/authors_2016_06.html}
%\and
%           Last address
          }

\abstract{%
The Auger Engineering Radio Array (AERA), at the Pierre Auger Observatory in Argentina, measures the radio emission of extensive air showers in the 30-80~MHz frequency range. AERA consists of more than 150 antenna stations distributed over 17 km$^2$. Together with the Auger surface detector, the fluorescence detector and the under-ground muon detector (AMIGA), AERA is able to measure cosmic rays with energies above 10$^{17}$~eV in a hybrid detection mode. AERA is optimized for the detection of air showers up to 60$^{\circ}$ zenith angle, however, using the reconstruction of horizontal air showers with the Auger surface array, very inclined showers can also be measured. In this contribution an analysis of the AERA data in the zenith angle range from 62$^{\circ}$ to 80$^{\circ}$ will be presented. CoREAS simulations predict radio emission footprints of several km$^2$ for horizontal air showers, which are now confirmed by AERA measurements. This can lead to radio-based composition measurements and energy determination of horizontal showers in the future and the radio detection of neutrino induced showers is possible.
}
\maketitle
\section{Horizontal Air Showers}
\begin{figure}[h]
\centering
\includegraphics[width=0.6\textwidth,clip]{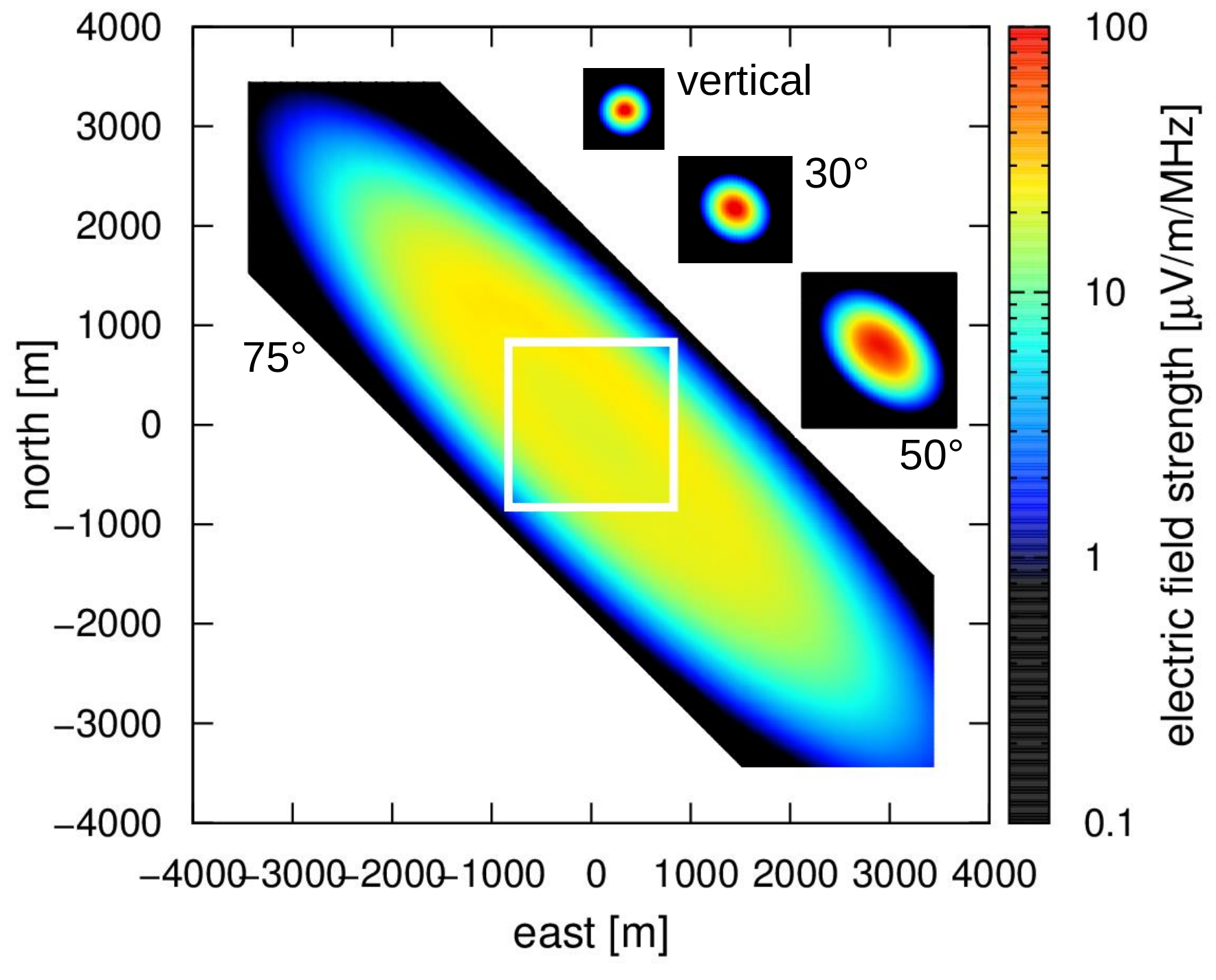}
\caption{Field strength of the radio emission as a function of ground position for CoREAS simulations of primary proton showers of 10$^{18}$ eV for different zenith angles. The white box marks the change of antenna distance from 40 m to 100 m in the simulation to save computing time.~\cite{theory, diss}}
\label{fig_coreas_sims}      
\end{figure}

\begin{figure}[h]
\centering
\includegraphics[width=\textwidth,clip]{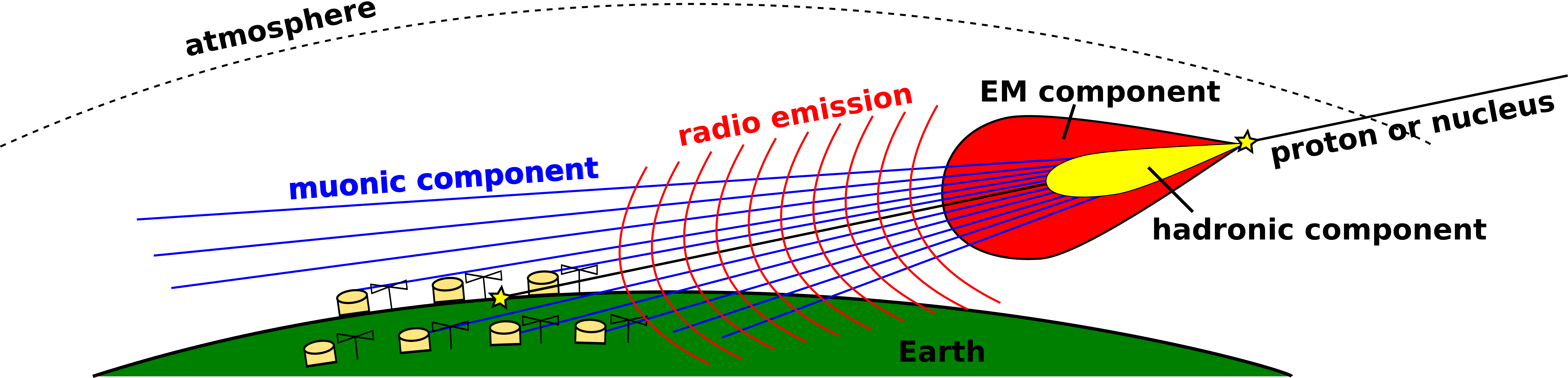}
\caption{Illustration of a horizontal air shower. The hadronic and electromagnetic component is absorbed and the muonic component and the radio signal can be measured on the ground.~\cite{arena_proc, diss}}
\label{fig_auger}       
\end{figure}

The radio emission of horizontal air showers covers a larger area on ground than the radio emission of vertical showers of the same energy and azimuth angle. CoREAS~\cite{coreas} simulations have predicted the large footprint of horizontal air showers of several km$^2$, which can be seen in figure~\ref{fig_coreas_sims}. This makes it possible to increase the distance between the antenna stations, as is done for the current phase of AERA with up to 750 m distance between the antenna stations. The Pierre Auger Observatory~\cite{auger} with its 3000 km$^2$ size provides the optimal conditions to detect horizontal air showers because the SD tanks are not flat like scintillators. The sketch of figure~\ref{fig_auger} illustrates that with the radio detection of extensive air showers the electromagnetic component of horizontal air showers can be accessed by detecting the radio emission even if the particle shower is absorbed in the atmosphere. The 1.5 km grid of the surface detector water-Cherenkov tanks triggers the externally triggered AERA~\cite{aera} antenna stations. The self-triggered stations of AERA are only sensitive up to 45$^{\circ}$ zenith angle and cannot be used for detecting horizontal air showers.\\

\section{Radio Detection of Horizontal Air Sowers with AERA}

\begin{figure}[h]
\centering
\includegraphics[width=0.35\textwidth,clip]{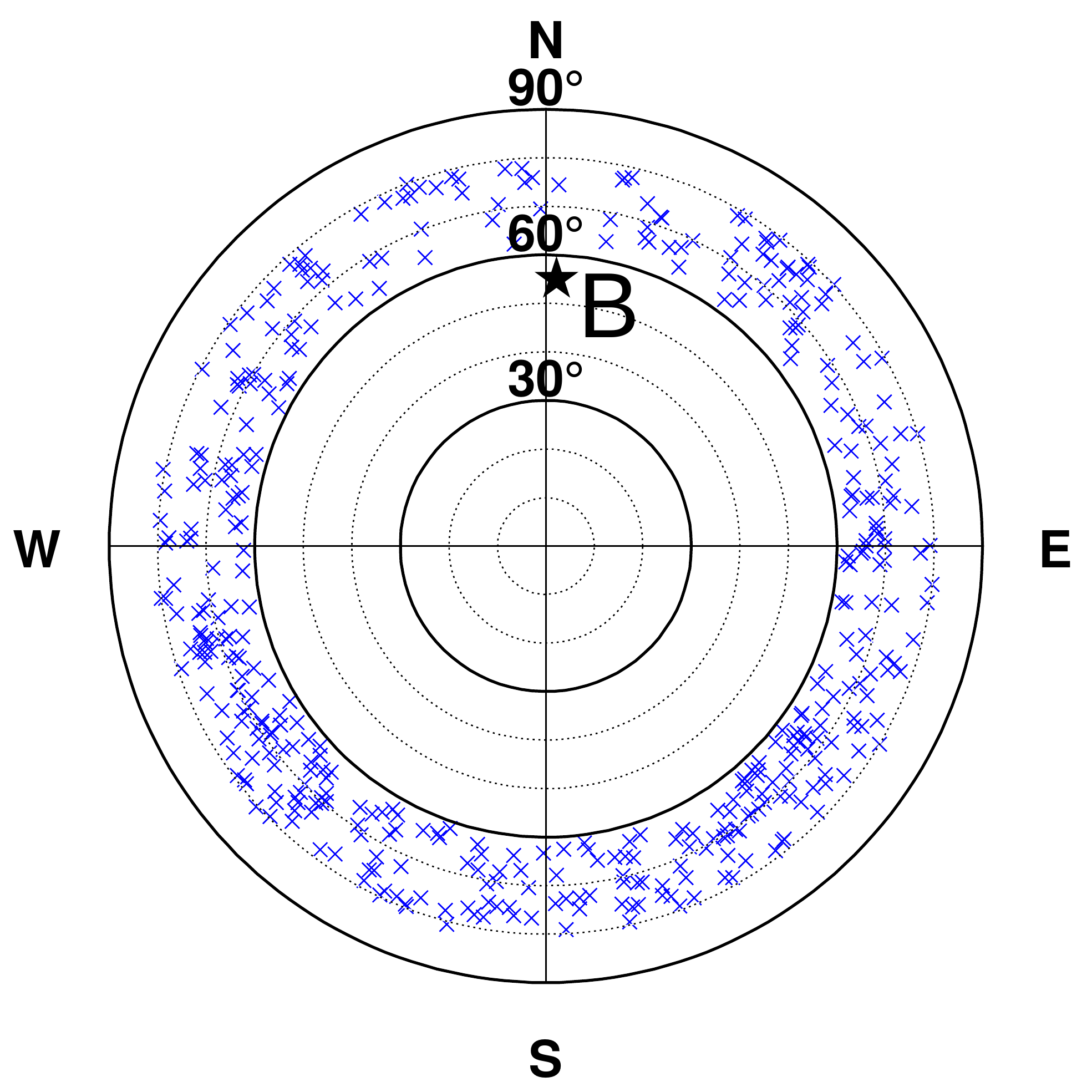}
\includegraphics[width=0.6\textwidth,clip]{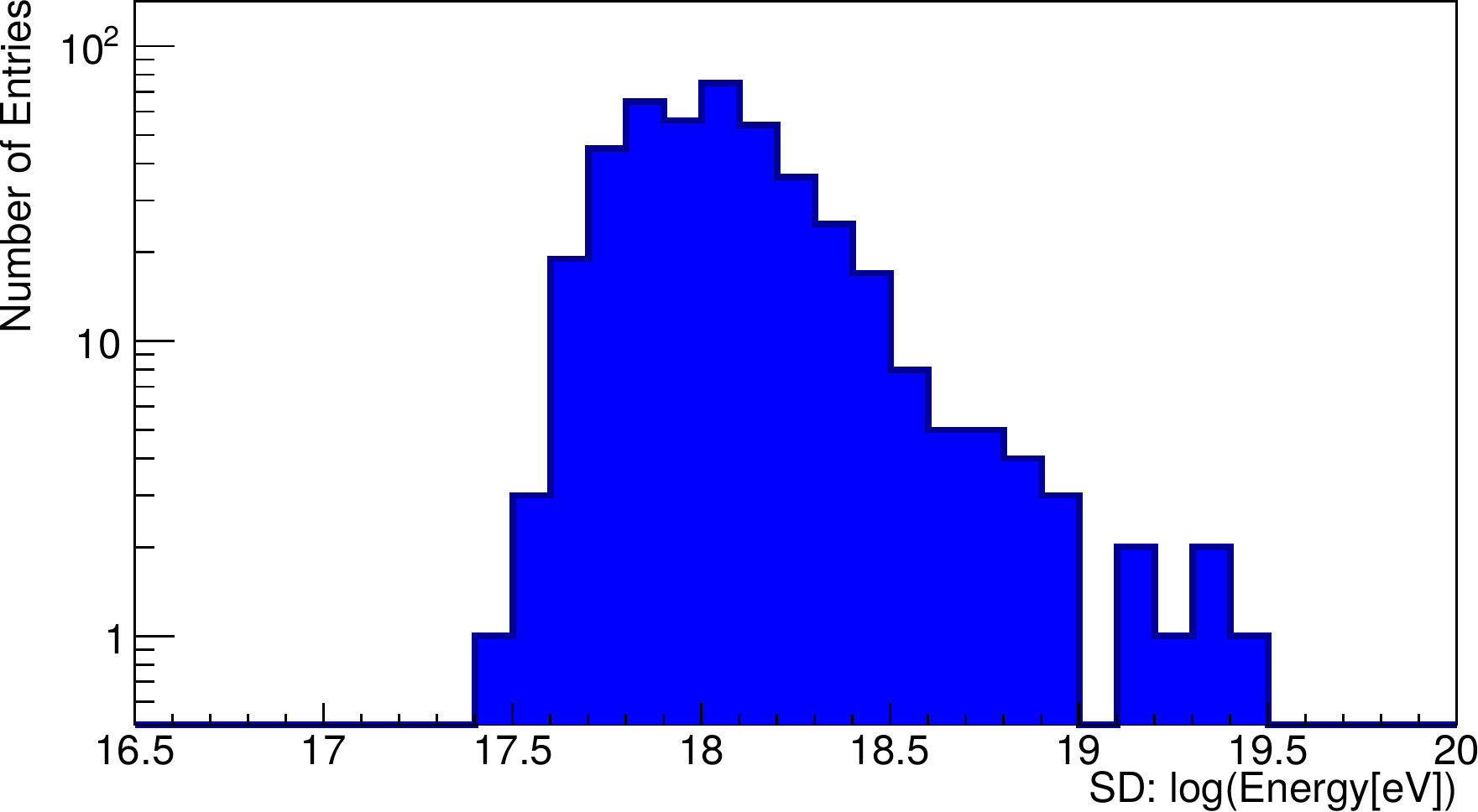}\\
\caption{The data sample and data selection of the externally triggered AERA events in the period from 1st of January 2012 to 15th of August 2015 results in 427 high quality events.}
\label{fig_events}       
\end{figure}

\begin{figure}[h]
\centering
\includegraphics[width=0.9\textwidth,clip]{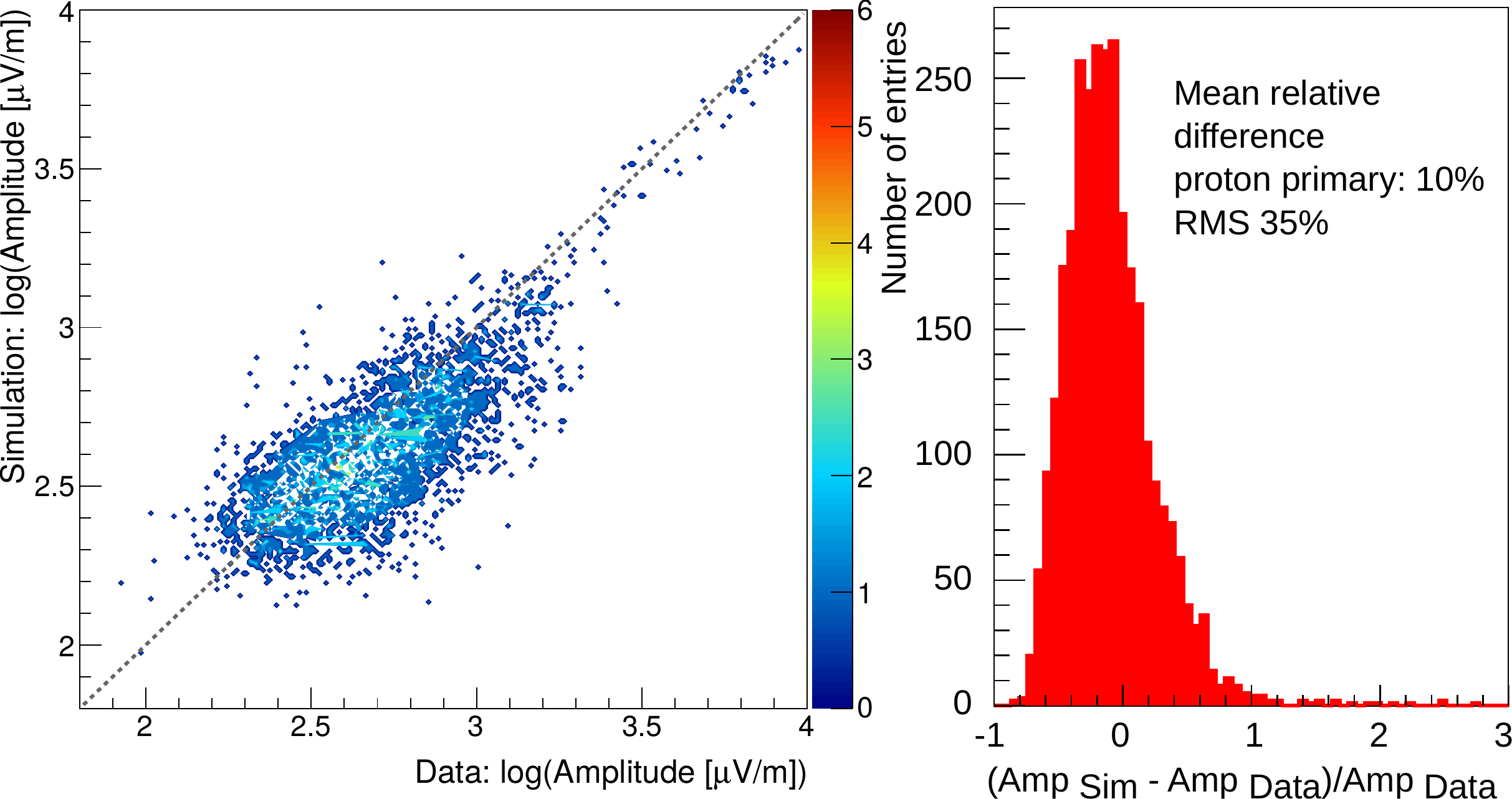}
\caption{Comparison of measured and simulated events. In 427 horizontal showers 3065 AERA stations are above signal-to-noise threshold. The chosen simulation code is CoREAS. The systematic shifts and uncertainties need to be further investigated.}
\label{fig_amp}      
\end{figure}

\begin{figure}[h]
\centering
\includegraphics[width=0.428\textwidth,clip]{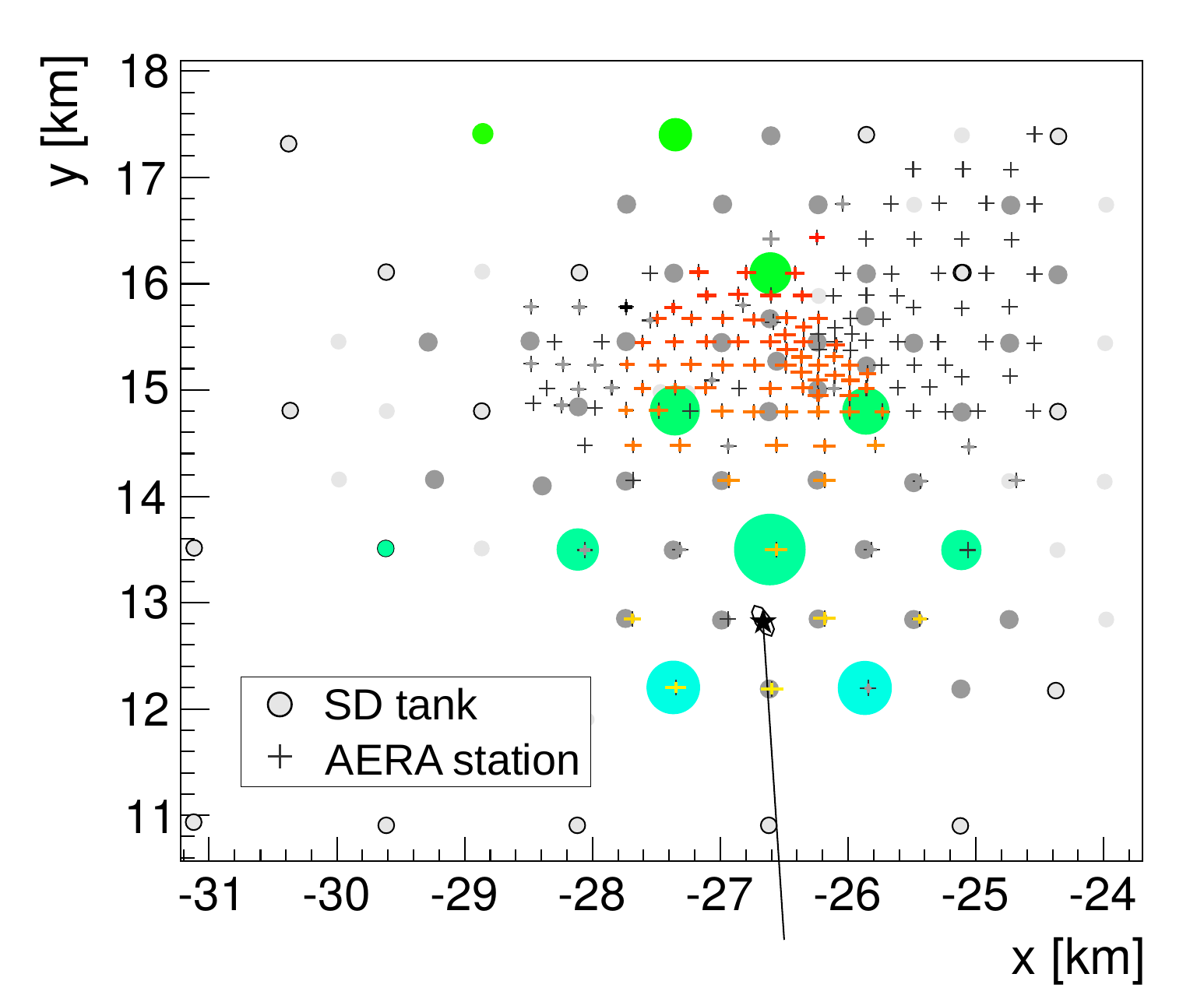}
\includegraphics[width=0.5\textwidth,clip]{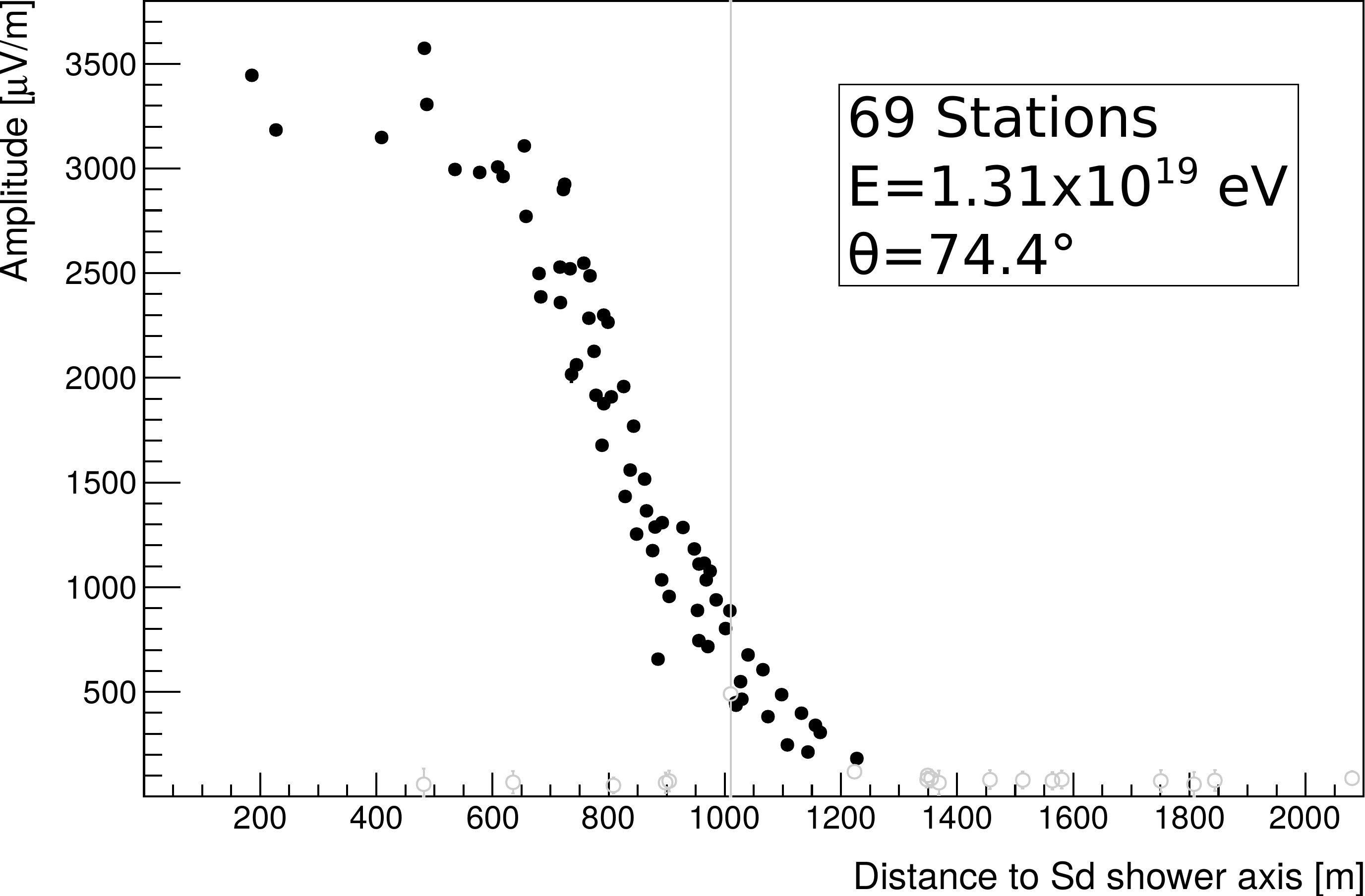}
\caption{Example event of a horizontal air shower with the current AERA set-up. The grid of the surface detector stations and antenna stations can be seen on the left and on the right the radio lateral distribution can be seen. The color scheme of the SD tanks and the AERA stations on the left indicate the timing information of the triggered tanks and stations and thus the arrival direction of the shower. The size of the circles indicates the signal strength.}
\label{example}      
\end{figure}

Data selection of surface-detector-triggered and reconstructed horizontal events of the AERA data sample in the period 1st of January 2012 to 15th of August 2015 results in 427 high quality events. The selection criterion of the zenith angle range from 62$^{\circ}$ to 80$^{\circ}$ is illustrated in the left plot of figure~\ref{fig_events} and the energy distribution of the detected events can be seen in the right plot of figure~\ref{fig_events}. Comparing the amplitude of each AERA station of each measured event with CoREAS simulations of proton primaries results in the left plot of figure~\ref{fig_amp}. The mean relative difference for CoREAS simulations with proton primaries is 10\% and the RMS is 35\%. These systematic shifts and uncertainties need to be further investigated, but the agreement illustrates that signals from horizontal air showers are basically well-described by the simulations.\\
The AERA horizontal air shower data sample contains 27 events with energy larger than 10 EeV, if no zenith angle cut is applied. These events have higher amplitudes and therefore allow deeper studies of the radio emission also in comparison with simulations. For half of the 427 horizontal showers the core is not contained in AERA and nevertheless a clear radio signal can be measured. One example event of the current AERA set-up can be seen in figure~\ref{example}.
This measured event reveals the potential of the radio detection of horizontal air showers. The event has an energy of 1.13$\times$10$^{19}$ eV, a zenith angle of 74.4$^{\circ}$ and contains 69 antenna stations above signal-to-noise threshold. The antenna stations with antenna distance of 750 m are contained and the detections of such kind of events is no problem on a 1.5 km grid.\\

\newpage
\section{Conclusion and Outlook}
The radio detection of horizontal air showers reveals great potential to be a comprehensive detection method to access the electromagnetic component of the extensive air shower and for horizontal air showers this is only possible with radio detection. The analysis of the AERA data sample results in 427 high quality events in the zenith angle range from 62$^{\circ}$ to 80$^{\circ}$. The large radio footprint covering several km$^2$ could be detected for the first time with the large-scale radio detector AERA. The radio signal of horizontal air showers is understood within the uncertainties. Sparse antenna arrays are well-suited for horizontal air shower measurements. Now the physics of the radio detection of horizontal air showers for composition measurements and energy determination is in reach and the detection of neutrino induced air showers possible, but more studies are needed in sensitivity and systematics.


\begin{thebibliography}{}
\bibitem{coreas}
T. Huege, M. Ludwig, C.W. James, Proc. ARENA 2012, Erlangen, Germany, AIP Conference Proc. 1535, 128, 2013.

\bibitem{auger}
A. Aab et al. (Pierre Auger Collaboration), Nucl. Instrum. Meth. A 798, 172 (2015), 1502.01323.

\bibitem{aera}
C. Glaser for the Pierre Auger Collaboration, this conference (2016)

\bibitem{theory}
T. Huege, Radio detection of cosmic ray air showers in the digital era, Phys.Rept. 620 (2016) 1-52.

\bibitem{diss}
O. Kambeitz, Radio Detection of Horizontal Extensive Air Showers, PhD thesis (2016), DOI(KIT):10.5445/IR/1000055758.

\bibitem{arena_proc}
O. Kambeitz, Radio Detection of Horizontal Extensive Air Showers with AERA, Proceedings of ARENA 2014, Anapolis, USA,   arXiv:1509.08289.

\end{thebibliography}
\end{document}